\pdfoutput=1
\documentclass[sigconf,natbib=true]{acmart}
\usepackage{import}

\newcommand{\paperKeywords}{recommendation systems, cold-start, out-of-vocabulary, inductive}

\newcommand{\paperTitle}{Improving Out-of-Vocabulary Handling \\ in Recommendation Systems}
\newcommand{\isDraft}{0}

\usepackage{amsmath,amsfonts,bm}

\def\eqref#1{equation~\ref{#1}}

\def\1{\bm{1}}

\def\vv{{\bm{v}}}

\def\mA{{\bm{A}}}
\def\mB{{\bm{B}}}

\def\mI{{\bm{I}}}

\def\mR{{\bm{R}}}

\def\mU{{\bm{U}}}

\DeclareMathAlphabet{\mathsfit}{\encodingdefault}{\sfdefault}{m}{sl}
\SetMathAlphabet{\mathsfit}{bold}{\encodingdefault}{\sfdefault}{bx}{n}

\def\gI{{\mathcal{I}}}

\def\gR{{\mathcal{R}}}

\def\gU{{\mathcal{U}}}
\def\gV{{\mathcal{V}}}

\def\sZ{{\mathbb{Z}}}

\newcommand{\R}{\mathbb{R}}

\long\def\IfNoTokA #1\IfNoTokB \IfNoTokB {}
\long\def\IfNoTokB \IfNoTokC #1#2{#1}    
\long\def\IfNoTokC           #1#2{#2}%
\makeatletter
\newcommand*{\tp}{%
  {\mathpalette\@transpose{}}%
}
\newcommand*{\@transpose}[2]{%
  \raisebox{\depth}{$\m@th#1\intercal$}%
}
\makeatother

\long\def\IfNoTokens #1{\IfNoTokA\IfNoTokB #1\IfNoTokB\IfNoTokB\IfNoTokC }

\definecolor{aqua}{rgb}{0.0, 1.0, 1.0}

\newcommand{\wtodo}[1]{{\textcolor{blue}{\textbf{[\IfNoTokens{#1}{TODO}{TODO: #1}]}}}}

\newcommand{\dset}[1]{\texttt{#1}}

\newcommand{\cmark}{\ding{51}}%
\newcommand{\xmark}{\ding{55}}%

\newcommand{\yelp}{\dset{Yelp-2018}}
\newcommand{\lfm}{\dset{LastFM-artists}}
\newcommand{\ham}{\dset{H\&M}}
\newcommand{\snap}{\dset{Content}}

\newcommand{\arxivOnly}[1]{}

\newcommand{\cmean}{\operatorname{cmean}}

\newcommand{\users}{\gU}
\newcommand{\items}{\gI}
\newcommand{\inter}{\gR}
\newcommand{\intermat}{\mR}
\newcommand{\ufeat}{\mU}
\newcommand{\ifeat}{\mI}
\newcommand{\tusers}{\users_{\text{train}}}
\newcommand{\tinter}{\inter_{\text{train}}}
\newcommand{\eusers}{\users_{\text{eval}}}
\newcommand{\titems}{\items_{\text{train}}}
\newcommand{\eitem}{\items_{\text{eval}}}

\newcommand{\formatEmbedder}[1]{\texttt{#1}}
\newcommand{\zero}{\formatEmbedder{zero}}
\newcommand{\random}{\formatEmbedder{rand}}
\newcommand{\meane}{\formatEmbedder{mean}}
\newcommand{\rbuck}{\formatEmbedder{r-bucket}}

\newcommand{\lsh}{\formatEmbedder{m-lsh}}
\newcommand{\slsh}{\formatEmbedder{s-lsh}}
\newcommand{\dhe}{\formatEmbedder{dhe}}
\newcommand{\dnn}{\formatEmbedder{dnn}}
\newcommand{\fdhe}{\formatEmbedder{fdhe}}
\newcommand{\knn}{\formatEmbedder{knn}}

\newcommand{\fzero}{f_{\text{zero}}}
\newcommand{\fknn}{f_{\text{knn}}}
\newcommand{\fmean}{f_{\text{mean}}}
\newcommand{\frand}{f_{\text{rand}}}

\newcommand{\fmaskRatio}{\beta}
\newcommand{\oovSampleRatio}{\alpha}

\newcommand{\smalltonormalsize}{%
  \fontsize
    {\fpeval{(\f@size@small+\f@size@normalsize)/2}}
    {\fpeval{(\f@baselineskip@small+\f@baselineskip@normalsize)/2}}%
  \selectfont
}

\newcommand\blfootnote[1]{%
  \begingroup
  \renewcommand\thefootnote{}\footnote{#1}%
  \addtocounter{footnote}{-1}%
  \endgroup
}

\usepackage{duckuments}     %
\usepackage{amsmath,graphicx,amsthm}
\usepackage[ruled,linesnumbered]{algorithm2e}
\usepackage{algorithmic}
\usepackage{listings}
\usepackage{cuted}
\usepackage{paralist} %
\usepackage{xfrac}       %
\usepackage{color}
\usepackage{rotating}
\usepackage{float}
\usepackage{hyperref}
\usepackage[mathscr]{eucal} %
\usepackage{amsbsy} %
\usepackage{subcaption}
\usepackage{booktabs}       %
\usepackage{tcolorbox}
\usepackage{multirow}
\usepackage{todonotes}
\usepackage{epstopdf}
\usepackage{balance}
\usepackage{tablefootnote}
\usepackage{footnote}
\usepackage[symbol]{footmisc}
\usepackage{printlen}
\usepackage{empheq} %
\usepackage{moresize}
\usepackage{enumitem}
\usepackage{graphicx}
\usepackage[utf8]{inputenc} %
\usepackage{nicematrix}
\usepackage{wrapfig} %
\usepackage[capitalize,noabbrev,nameinlink]{cleveref}
\setlist{nolistsep}
\usepackage{pifont} %
\usepackage{xfp}
\usepackage{anyfontsize}
\usepackage{layouts}
\usepackage{microtype}      %

\ifnum\pdfstrcmp{\jobname}{output}=0 %
    \usepackage[outputdir=../,frozencache]{minted}      %
\else %
    \usepackage[frozencache]{minted}
\fi

\DeclareCaptionType{copyrightbox}
\usepackage{url}

\renewcommand{\algorithmicrequire}{\textbf{Input:}}
\renewcommand{\algorithmicensure}{\textbf{Output:}}
\newcounter{ALC@tempcntr}%

\hypersetup{
colorlinks=true,
linkcolor=red,
anchorcolor=blue,
citecolor=brown
}

\if \isDraft 1%
\paperwidth=\dimexpr \paperwidth + 8cm\relax
\oddsidemargin=\dimexpr\oddsidemargin + 4cm\relax
\evensidemargin=\dimexpr\evensidemargin + 4cm\relax
\fi%

\copyrightyear{2023}
\acmYear{2023}
\setcopyright{rightsretained}

\makeatother

\begin{document}

\title{\paperTitle{}}

\author{William Shiao$^{1,*}$, Mingxuan Ju$^{2,3}$, Zhichun Guo$^{2}$, Xin Chen$^{3}$, Evangelos Papalexakis$^{1}$, Tong Zhao$^{3}$, Neil Shah$^{3}$, Yozen Liu$^{3}$}
\affiliation{
  \institution{$^1$University of California, Riverside, $^2$University of Notre Dame, $^3$Snap Inc. \\
  \{wshia002@,epapalex@cs.\}ucr.edu;\; \{mju2,zguo5\}@nd.edu;\;
  \{mju,xin.chen,tong,nshah,yliu2\}@snap.com
  }
  \country{}
}

\renewcommand{\shortauthors}{William Shiao et al.} %

\begin{abstract}
  Recommendation systems (RS) are an increasingly relevant area for both academic and industry researchers, given their widespread impact on the daily online experiences of billions of users. 
One common issue in real RS is the \emph{cold-start} problem, where users and items may not contain enough information to produce high-quality recommendations. This work focuses on a complementary problem: recommending new users and items unseen (out-of-vocabulary, or OOV) at training time. %
This setting is known as the \textit{inductive} 
setting and is especially problematic for factorization-based models, that rely on encoding only those users/items (and, more generally, other sparse features) seen at training time with fixed parameter vectors. However, despite its practical significance, handling OOV values is often an afterthought in many academic works due to a predominant focus on \emph{transductive} evaluation, where all categorical values for sparse features are observed at training time. %
As a result, existing solutions applied in practice are often na\"\i{}ve, such as assigning OOV users/items to random buckets.
In this work, we tackle this problem and propose approaches that better leverage available user/item features to improve OOV handling at the embedding table level. We discuss general-purpose plug-and-play approaches which are easily applicable to most RS models and improve inductive performance without negatively impacting transductive model performance.
We extensively evaluate 9 OOV embedding methods on 5 models across 4 datasets (spanning different domains). One of these datasets is a 
proprietary production dataset from a prominent RS employed by a large social platform serving hundreds of millions of daily active users.
In our experiments, we find that several proposed methods that exploit feature similarity using LSH consistently outperform alternatives on a majority of model-dataset combinations, with the best method showing a mean improvement of $3.74\%$ over the industry standard baseline in inductive performance. We release our code and hope our work helps practitioners make more informed decisions when handling OOV for their RS and further inspires academic research into improving OOV support in RS\@.

  \vspace{-0.2in}
  \blfootnote{*Work done during an internship at Snap Inc.}
\end{abstract}

\settopmatter{printfolios=true}
\maketitle

\section{Introduction}%
\label{sec:intro}
\begin{figure}[!ht]
    \centering
    \includegraphics[width=0.7\columnwidth]{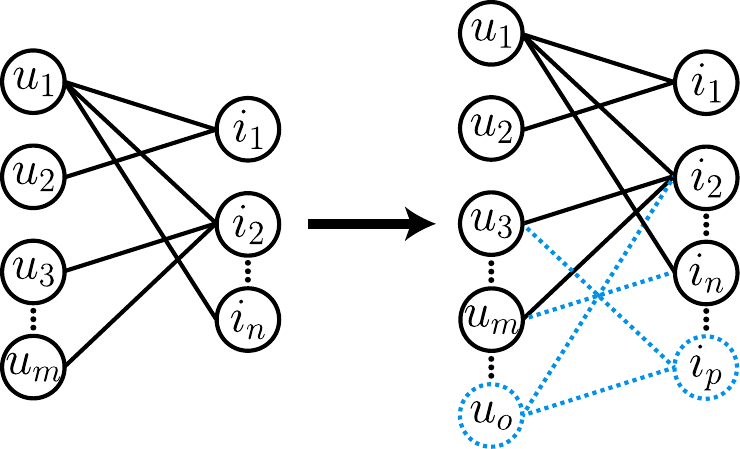}
    \caption{
    Comparison between transductive (left) and inductive (right) settings. 
    In the transductive setting, RS are evaluated on interactions between users and items observed during training time (i.e., bold links). Whereas in the inductive setting, besides transductive interactions, RS are also evaluated on interactions related to users and items unseen during the training (i.e., both bold and dash links).
    }
    \label{fig:ind_setting_overview}
    \Description[Overview of inductive setting]{An overview of the inductive setting, showing examples of inductive and transductive nodes.}
\end{figure}
Recommendation systems (RS) suggest items to users and have found wide adoption across a variety of domains. For example, they have been used to recommend advertisements~\cite{dcn,dcnv2}, movies~\cite{movielens}, friends~\cite{sankar2021graph,shi2023embedding}, and products~\cite{dong2016combining,linden2003amazon} to users. These methods are studied in both academia and industry, but many aspects often differ between academic and industrial recommendation systems~\cite{schnabel2022situating}. One such difference is their evaluation methodology. 

RS research in academia primarily focuses on the \emph{transductive} setting~\cite{wang2022towards,sun2023challenges}, where a portion of interactions are masked out for validation and testing. Such a setting assumes that all users and items in the dataset are seen during training. However, in industrial RS environments, there is often a constant influx of new, or out-of-vocabulary (OOV), users and items that were not seen at training time, i.e., the \textit{inductive setting} which correspond to new users and/or items showing up at validation and testing. Almost all production models are deployed to be utilized in an (at least partially) inductive setting, %
but a recent survey~\cite{schnabel2022situating} found that only about 10\%\footnote{based on an estimate from Figure 1 of~\cite{schnabel2022situating}.} of 88 recent RS papers evaluated their models in the fully inductive setting, in which OOV users and items are considered.
Similarly, existing state-of-the-art models~\cite{cai2023lightgcl,yu2023xsimgcl,dcnv2} also use embedding tables for sparse ID features,
which face similar issues when encountering values unseen at training time since a model would not have an existing row in the embedding table for unseen values.

\begin{figure}[t!]
    \centering
    \includegraphics[width=\columnwidth]{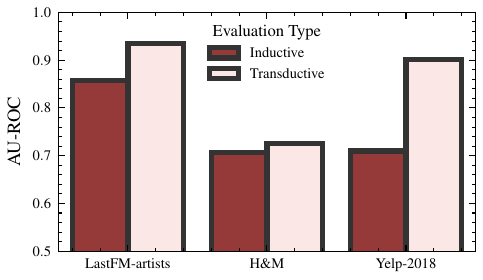}
    \vspace{-0.3in}
    \caption{Comparison of inductive vs transductive performance with Wide \& Deep models, where OOV (inductive) values are handled with trained random buckets. We see a clear gap in inductive performance vs transductive performance, showing the importance of properly handling OOV values.}%
    \label{fig:ind_trans_comparison}
    \vspace{-0.20in}
\end{figure}

While handling the inductive setting, industrial practitioners often use primitive methods such as random hashing to a fixed number of OOV buckets whose values are updated during training\footnote{\label{foot:examples}Some examples of industry usage are \url{https://engineering.linkedin.com/blog/2023/enhancing-homepage-feed-relevance-by-harnessing-the-power-of-lar}, \url{https://blog.taboola.com/preparing-for-the-unexpected/} and \href{https://github.com/bytedance/monolith/blob/135c491a52b151772b976af989d8bc938c44d210/monolith/core/feature.py\#L436}{in the Monolith source code}~\cite{liu2022monolith}.}. 
Such random hashing on OOV values intuitively can unlock the inductive capability for almost any transductive RS without affecting transductive performance and have been incorporated into industrial RS infrastructures as default options~\cite{tensorflow2015}. 
However, while simple to implement, these primitive methods can easily map two very different OOV users/items to the same embedding bucket, which is known as embedding collision and can greatly affect the recommendation performance~\cite{liu2022monolith,zhang2020model}. In \cref{fig:ind_trans_comparison}, we show that with a primitive method like assigning OOV values to random buckets, there exists a clear gap between inductive and transductive performance for all datasets. This demonstrates the importance of properly handling OOV values and leads us to the following question: \textbf{Can we re-imagine how we handle OOV users and items to improve the inductive capability of RS\@?}

While many methods in literature could potentially be used to solve this problem, we constrain our search to methods that meet criteria important to industry practitioners:

\begin{itemize}[leftmargin=*]
    \item \textit{Efficient}: the OOV embedding method should run in sub-linear time with respect to the total number of users/items.
    \item \textit{Maintains Transductive Performance}: active users and popular items are often the platform's main income sources. Hence, the OOV embedding method should not sacrifice the base model's performance on non-OOV items.
    \item \textit{Model-Agnostic}: the OOV embedding method should be applicable to RS with different model architectures. 
\end{itemize}

Given the above criteria, this work explores existing and proposes new OOV embedding methods. These methods range from simply using a zero vector to feature-similarity-based methods.
In our experiments, we thoroughly evaluate nine different OOV embedding methods (detailed in section \ref{sec:embedders}) to provide a broad empirical understanding of the performance of different OOV strategies. 
Among the 9 OOV methods, inspired by feature-based cold-start work~\cite{lam2008coldstart}, we propose using several feature-based methods, which utilize feature information to compensate for OOV values.
In particular, we propose two locality-sensitive hashing (LSH)~\cite{gionis1999similarity} based methods that exploit feature-similarity consistently outperform other feature or non-feature-based methods in most models-dataset combinations, with the best method showing a mean improvement of $3.74\%$ over the industry-standard random bucket assignment method.

To properly evaluate OOV methods under inductive settings, we also create appropriate inductive datasets, as existing public datasets are (1) transductive and (2) lack user/item features. Such limitations directly contradict the setting faced in industrial recommendation systems, where we usually have rich feature information for both users and items and many OOV values. Therefore, we augment three existing open-source datasets and perform a time-based split such that unseen items naturally appear during evaluation.
Furthermore, we also created a proprietary industrial dataset from a large social media company containing rich feature information to evaluate OOV methods properly under real applications. 

Our contributions can be summarized as the following,

\begin{itemize}[leftmargin=*]
    \item To the best of our knowledge, our work is the first to provide a comprehensive empirical understanding of the performances of various OOV methods for RS.
    \item We demonstrate that a class of proposed feature-aware, locality-sensitive hashing-based OOV embedders that exploit feature-similarity consistently outperform existing approaches in inductive performance.
    \item We provide realistic inductive datasets by augmenting and splitting three open-source datasets, enabling experiments on inductive performance and OOV methods of RS, which will be publicly available upon the release of this manuscript.
    \item We will open-source our evaluation framework, a major extension of the popular RecBole~\cite{zhao2021recbole} RS library that adds inductive and OOV support to encourage future research in this area.
\end{itemize}%
\section{Preliminaries and Related Work}
\label{sec:preliminaries}

In this section, we formally define OOV values and the user/item recommendation system problem. 
We detail the difference between the two classes of RS model setups that we study, delineated by the use of contextual features: context-free vs. context-aware models since OOV handling behaves differently for each class of models.

\begin{figure*}[!htp]
    \centering
    \includegraphics[width=0.8\linewidth]{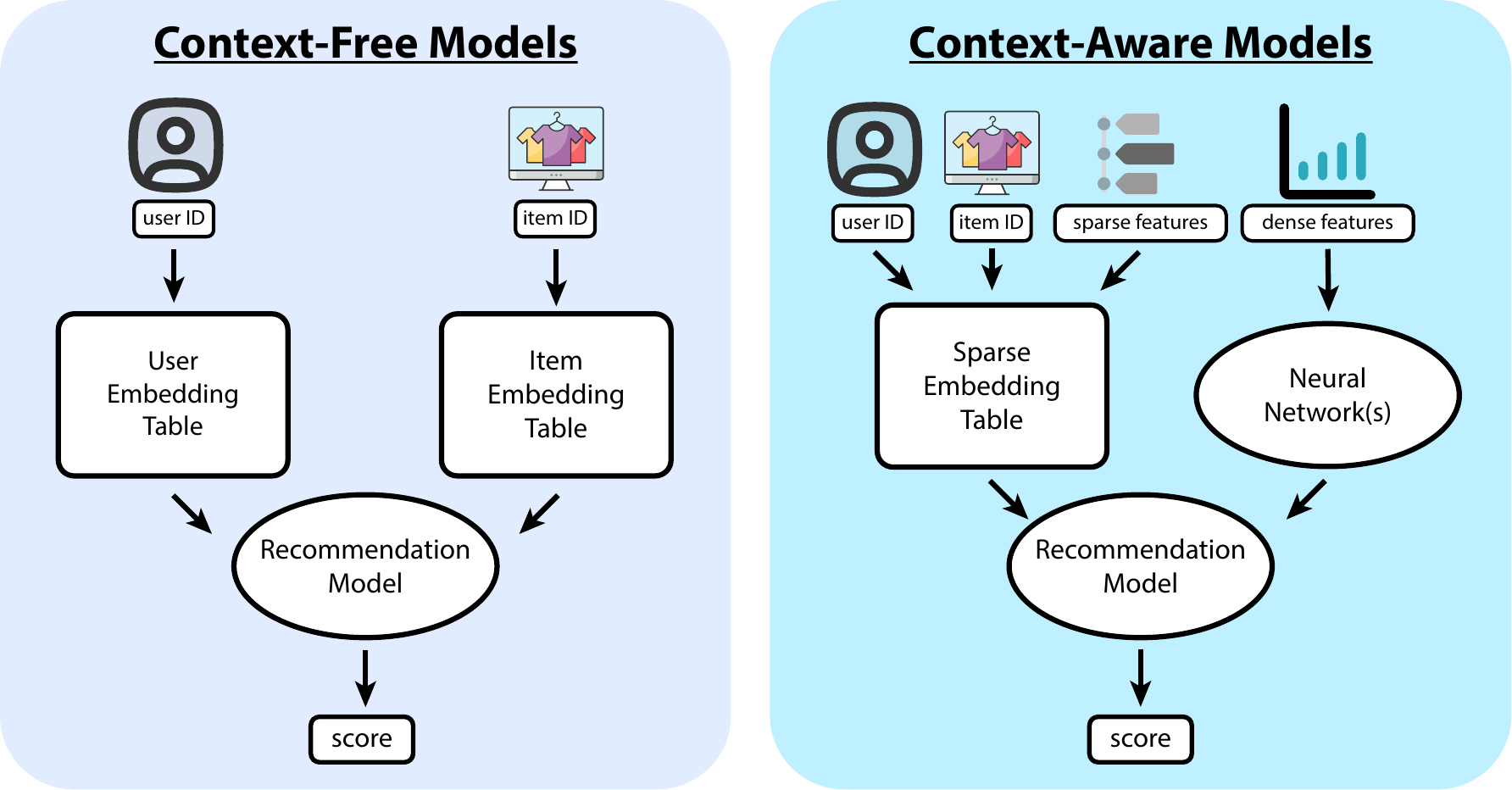}
    \caption{Typical structure of context-aware and context-free recommendation models.}%
    \label{fig:recsys_overview}
    \Description[Typical structure RS models]{Typical structure of context-aware and context-free recommendation models.}
\end{figure*}

\paragraph{Notation} We denote the set of users as $\users$ and the set of items as $\items$. We denote the set of interactions as $\inter \subseteq \users \times \items$. 
For flexibility, let $\intermat$ be the interaction matrix such that $\intermat_{u,i} = 1 \iff (u, i) \in \inter$. Let $m = |\users|$ and $n = |\items|$ be the number of users and items, respectively. Let $\ufeat \in \R^{m \times d}$ and $\ifeat \in \R^{n \times d}$ be the user/item feature matrices. We assume\footnote{We assume users/item features to have the same dimension $d$ for simplicity, but this can be enforced in practice with a projection layer if user/item features have different dimensions $d_u$ and $d_i$ respectively.} that both feature matrices are of dimension $d$. For a given user $u \in \users$, we have the associated contextual features $\ufeat_u$. Similarly, for a given item $i \in \items$, we have the associated features $\ifeat_i$. $\cmean(\cdot):\mathbb{R}^{n \times d} \rightarrow \mathbb{R}^{d}$ is the column-wise mean of a matrix.

We split the set of users and items based on a time $t$. All users/items appearing before time $t$ are considered a part of the training set, users $\tusers \subseteq \users$ and items $\titems \subseteq \items$. The set of training interactions $\tinter \subseteq \inter$ is also similarly created.

\paragraph{OOV Values} We consider a value Out-Of-Vocabulary (OOV) if it is a categorical value that does not exist at training time but appears at inference time. Formally, a user $u$ is OOV if and only if $u \not\in \tusers \land u \in \users$ and an item $i$ is OOV if and only if $i \not\in \titems \land i \in \items$. We abbreviate non-OOV values as IV (In-Vocabulary).

\paragraph{Transductive vs. Inductive Settings} In the transductive setting, RS models are evaluated on interactions between users and items that are observed during the model training (i.e., $\eusers \subseteq \tusers$ and $\eitem \subseteq \titems$). 
Whereas in the inductive setting, besides transductive interactions, RS models are also evaluated on interactions between users and items that do not appear during the model training (i.e., $\eusers \cup \tusers \neq \tusers$ and $\eitem \cup \titems \neq \titems$).

\subsection{Context-Free Models}%
\label{subsec:ctx-free}
Context-free models are the ones that do not use any additional feature information other than the IDs of users or items. They are also known as latent factor models~\cite{shah2023survey} and are typically based on matrix factorization (MF)~\cite{koren2009matrix,koren2008factorization}, with the goal of approximating the training interaction matrix $\intermat_{\text{train}} \in \mathbb{R}^{m \times n}$. Typically, they factor $\intermat_{\text{train}}$ into two matrices $\mA  \in \mathbb{R}^{m \times d}$ and $\mB  \in \mathbb{R}^{n \times d}$ such that $\intermat_{\text{train}} \approx \mA \mB^\top$. The rows of $\mA$ and $\mB$ are the user and item embeddings, respectively.

These embeddings can be learned in a variety of ways. For example, the Non-negative Matrix Factorization (NMF)~\cite{nmf} of the interaction matrix can be computed via non-negative least squares or gradient descent. In this work, we focus on two popular context-free models: Bayesian Personalized Ranking (BPR)~\cite{rendle2012bpr} and DirectAU~\cite{wang2022towards}. 

BPR is a pairwise ranking model that learns user and item embeddings by maximizing the likelihood of observed interactions.
\begin{equation}
    \label{eq:bpr_loss}
    \mathcal{L}_{\text{BPR}} = \frac{1}{|\inter|} \sum_{(u,i) \in \inter} -\log \left(\sigma(\mA_u \cdot \mB_i^\top - \mA_u \cdot \mB_{i'}^\top)\right),
\end{equation}
where $i'$ is a randomly sampled item from $\items$ such that $(u, i') \not\in \inter$ and $\sigma(\cdot)$ is the sigmoid function.
In \cref{eq:bpr_loss}, $\mA_u \cdot \mB_i^\top$ could be regarded as the dot-product similarity between user $u$ and item $i$.
One recent work~\citep{he2017neural} replaces the dot-product similarity by an MLP to infer a similarity score. 
Unlike BPR that utilizes negative sampling (i.e., $\mA_u \cdot \mB_{i'}^\top$ in \cref{eq:bpr_loss}) for training, DirectAU~\cite{wang2022towards} is a loss function that instead directly optimizes for alignment and uniformity --- factors that have been shown to be important for representation quality~\cite{wang2020understanding}. It is worth noting that these are often used as retrieval models in production~\citep{shi2023embedding}, 
which is why we evaluate them as such in our experiments.

\subsection{Context-Aware Models}%
\label{subsec:ctx-aware}

Context-aware models utilize complimentary contextual features in addition to the user or item IDs. They are often based on the two-tower architecture~\cite{huang2013learning}, where each tower is responsible for embedding the user and item features, respectively. The two towers output embeddings of the same dimensionality, allowing them to be directly compared to produce a score for each user-item pair.

However, these models are very dependent on the quality of the input contextual features. In production, practitioners often produce cross-features~\cite{cheng2016wide} that capture the interactions between features. As such, we focus on 3 context-aware models that incorporate these cross-features: Wide \& Deep~\cite{cheng2016wide}, eXtreme Deep Factorization Machine (xDeepFM)~\cite{lian2018xdeepfm}, and Deep \& Cross Networks V2 (DCN-V2)~\cite{dcnv2}. We focus on these models as they are three of the most popular context-aware models in practice. The models are often used during the ranking or re-ranking stage in production pipelines, hence we evaluate them using ranking metrics in our experiments.

The features used in context-aware models are typically categorized into two types: sparse and dense. Sparse features are categorical features that are typically one-hot or multi-hot encoded. Dense features are continuous features. For example, in the case of social media content recommendation, a user's country could be a sparse feature and their mean daily app usage could be a dense feature. Sparse features are typically embedded using an embedding table 
where each row represents the embedding for that feature's ID\@. These tables are typically randomly initialized and gradually updated during training. Dense features are typically either unmodified or passed through neural network layers. In this work, we focus primarily on the handling of OOV values in sparse features --- specifically, the user/item IDs\@, which are most likely to be OOV in production settings.
\section{Towards a General OOV Embedder}%
\label{sec:embedders}

The motivation for this work stems from how OOV users/items are typically handled in real-world production settings. In practice, OOV users/items are often assigned to a random bucket within which all values share the same embedding or are simply assigned completely random embeddings%
\footref{foot:examples}%
. This clearly results in poor performance for any pure ID-based models (e.g., factorization-based) that rely on stored embeddings for users/items seen at training time. However, even for models that use features, this can still result in poor performance since poorly-assigned embeddings simply add additional noise. For example, random bucket assignment for OOV users means that two OOV users have the same chance to share an embedding, regardless of how similar/different they are. 

Since our goal is to improve OOV support for most general recommendation systems, regardless of specific model architecture, we limit the scope of our modifications to a component that is used in almost all production recommendation systems: the embedding table. In this work, we focus primarily on OOV support for unseen user/item IDs, but the same ideas can also be easily extended to improve support for unseen categorical values in other features. 
This leads to the following formal definition of an OOV embedder:

\paragraph{OOV Embedders} A user OOV embedder $f_{\text{user}} : \users \setminus \tusers \rightarrow \R^d$ maps an OOV user to a real-valued embedding. An item OOV embedder does the same: $f_{\text{item}} : \items \setminus \titems \rightarrow \R^d$.

For the sake of simplicity, we describe all the following OOV embedders in terms of OOV users, but we use them for both OOV users and items during evaluation. They can be easily converted to item OOV embedders by substituting the appropriate variables.

\begin{table*}[h!]
	\centering
	\begin{tabular}{r|cccccccccc}
		\toprule
		\textbf{Embedder}                          & \zero{}  & \meane{} & \random{} & \rbuck{} & \knn{}                                                                                                                                                                          & \dhe{}      & \fdhe{}     & \dnn{}      & \lsh{}   & \slsh{}  \\ 
		\midrule
		Requires training                          & \xmark{} & \xmark{} & \xmark{}  & \cmark{} & \xmark{}                                                                                                                                                                        & \cmark{}    & \cmark{}    & \cmark{}    & \cmark{} & \cmark{} \\
		Uses user/item ID                          & \xmark{} & \xmark{} & \cmark{}  & \cmark{} & \xmark{}                                                                                                                                                                        & \cmark{}    & \cmark{}    & \xmark{}    & \xmark{} & \xmark{} \\
		Uses trainble OOV buckets                           & \xmark{} & \xmark{} & \xmark{}  & \cmark{} & \xmark{}                                                                                                                                                                        & \xmark{}    & \xmark{}    & \xmark{}    & \cmark{} & \cmark{} \\
		Uses features                              & \xmark{} & \xmark{} & \xmark{}  & \xmark{} & \cmark{}                                                                                                                                                                        & \xmark{}    & \cmark{}    & \cmark{}    & \cmark{} & \cmark{} \\
		Same features $\rightarrow$ same embedding & \xmark{} & \cmark{} & \xmark{}  & \xmark{} & \cmark{}                                                                                                                                                                        & \xmark{}    & \xmark{}    & \cmark{}    & \cmark{} & \cmark{} \\
		Requires pre-processing                    & \xmark{} & \cmark{} & \xmark{}  & \xmark{} & \cmark{}                                                                                                                                                                        & \xmark{}    & \xmark{}    & \xmark{}    & \xmark{} & \xmark{} \\
		Complexity                                 & $O(1)$   & $O(1)$   & $O(1)$    & $O(1)$   & <$O(n)$\tablefootnote[3]{The exact complexity here is difficult to compute since we rely on approximate nearest neighbor search~\cite{johnson2019billion,guo2020accelerating}.} & $O(\theta)$ & $O(\theta)$ & $O(\theta)$ & $O(b)$   & $O(b)$   \\
		Potential unique embeddings                & 1        & 1        & 1      & $b$   & $>n$                                                                                                                                                                            & $>n$        & $>n$        & $>n$        & $>n$     & $b$      \\
		\bottomrule
	\end{tabular}
	\caption{Comparison of the different OOV embedders evaluated in this work. For applicable methods, $\theta$ refers to the number of parameters in the neural network, $b$ refers to the number of buckets, and $n$ is the number of input items.\ \textit{Features} refer to non-ID features. We assume the embedding dimensionality is constant for the complexity analysis.}%
	\label{table:embedder_comparison}
\end{table*}%

\begin{figure}[!h]
    \centering
    \includegraphics[width=\columnwidth]{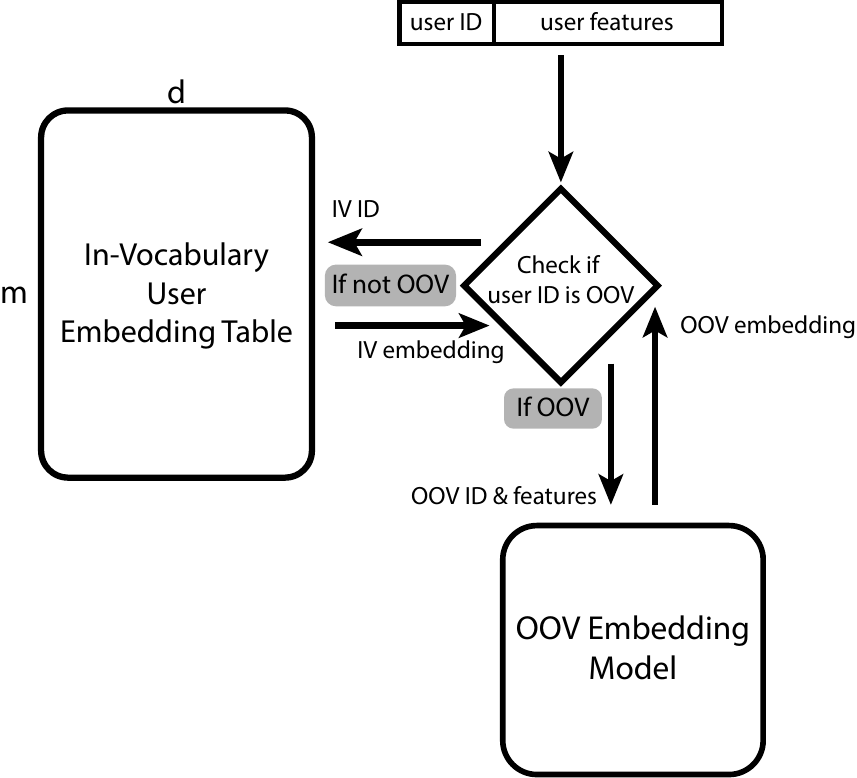}
    \caption{How IV/OOV user IDs are handled under our framework. Item IDs are handled the same way.}%
    \label{fig:oov_overview}
\end{figure}

\subsection{Heuristic-based Embedders}%
\label{subsec:untrained_embedders}

In this work, we first introduce several heuristic-based OOV embedding models that do not require additional trainable parameters. These are straightforward to apply in practice due to their speed and ease of implementation.

\paragraph{Zero Embedder} \zero{} simply uses the zero vector for all OOV inputs. This is a simple solution sometimes used in Natural Language Processing (NLP) for OOV words~\cite{lochter2020deep,mielke2021between}. Formally, $\fzero (\cdot) = {\{0\}}^c$. For context-free models, all new users will randomly select items (we ensure that items with the same score will be randomly selected without bias towards their ID). For context-aware models, all predictions will depend entirely on a user's/item's contextual information.

\paragraph{Mean Embedder} \meane{} uses the column-wise mean of the embedding matrix for all OOV IDs. Note that users and items use their respective means. Formally, for users, $\fmean (\cdot)= \cmean(\ufeat)$. For context-free models, this means that the RS model will recommend the same popular items to all new users and that all new items will have the same probability of being recommended.

\paragraph{Fixed Random Embedder} \random{} returns a random floating point vector for all OOV IDs. There are $b$ fixed random vectors for each ID type (e.g., user ID, item ID). This ensures that the model's output is deterministic for users/items. Formally, for a set of random vectors $\gV = \{ \vv \in \R^d \}$, we have $\frand (\cdot) = \gV_{g(u)}$ where $g$ is a random hash function\footnote{\label{foot:hash_func}We use the three-round integer hash function from \url{https://github.com/skeeto/hash-prospector}.} $\sZ \rightarrow \{ 1, 2, \ldots{} b \}$. This approach is similar to generating a random vector, except that (a) the output for a given ID is deterministic, and (b) the maximum amount of memory used is bounded by $b$.

\paragraph{KNN Embedder} \knn{} returns the mean of the $k$ nearest neighbors of a given point, as measured by the inner product of the features. Formally, for a user $u$, we have $\fknn (u) = \frac{1}{k} \sum_{a \in \textsc{K-Nearest}(u)} \ufeat_a$. With $k=2$, this is similar to the double-hashing performed by \citet{zhang2020model}, except that we use feature similarity instead of random hashing to select rows. In order to meet the efficiency criteria mentioned in \cref{sec:intro}, we use approximate nearest neighbor search through libraries like FAISS~\cite{johnson2019billion} and SCaNN~\cite{guo2020accelerating}. Each training ID's $k$-nearest neighbors can optionally be pre-computed and stored to prevent additional overhead during training.

\subsection{Learning-based Embedders}%
\label{subsec:trained_embedders}

We also consider a set of trained embedders that are optimized during the training of the base model. As mentioned in \cref{subsec:training}, we freeze the non-OOV parameters of the base model to avoid affecting its transductive performance. Some of these methods use an embedding table with $b$ rows, where each row corresponds to an OOV \textit{bucket}. This value can be tuned depending on the expected number of OOV values.
In the following paragraphs, we introduce several different learning-based OOV embedders. We describe how these embedders are optimized in \cref{subsec:training}.

\paragraph{Random Buckets} \rbuck{} randomly assigns an embedding (denoted as a bucket) to a given OOV ID\@. This mapping is done with a deterministic hash function\footref{foot:hash_func} to ensure that the bucket mappings remain consistent. The chance of any bucket being selected is uniform. Given $o$ OOV IDs and $b$ buckets, each bucket's expected number of OOV IDs is $\sfrac{o}{b}$. This is similar to \random{}, except that the values in each bucket are optimized during training. This is TensorFlow's~\cite{tensorflow2015} default approach for handling OOV values. PyTorch-style pseudocode for this approach can be found in \cref{algo:random_embedder}.

\paragraph{DHE} Deep Hash Embedding (DHE)~\cite{dhe} substitutes a deep neural network for the embedding table. To ensure determinism for a given ID, they first compute many hashes on that ID and use those as inputs to the neural network. We use SipHash~\cite{siphash} with different key values as the hash functions for our implementation. DHE was originally created as a drop-in replacement for the main embedding table in context-free methods, but we use it as an OOV embedded (only on OOV IDs) since it naturally works in this case.

\paragraph{F-DHE} \citet{dhe} mentions that DHE can also incorporate user features\@. \fdhe{} uses the concatenation of the user/item feature vector with the hash inputs (as with DHE) for the input to a deep neural network\@. This incorporates user/item features into the OOV embedding. However, compared to \dnn{}, it also assigns a unique embedding for each user/item ID, even if they share the same features.

\paragraph{DNN} \dnn{} is a simple feed-forward deep neural network that takes in the user/item features as input and outputs a real-valued vector. This embedder can be viewed as a modification to \fdhe{} that omits the hash-related features. As a result, users/items with the same features will share the same embedding.

\paragraph{Mean LSH} \lsh{} is a locality-sensitive hashing (LSH)~\cite{lsh} based OOV embedder. It uses a random projection matrix to map a user/item ID to a binary vector. It then uses this binary vector to index into the OOV embedding table and returns the column-wise mean of the rows where the binary vector is 1. This helps ensure that similar users/items have similar embeddings, even if their LSH vector is not exactly the same. The projection matrix remains constant, but the OOV embedding table values are updated during training. PyTorch-style pseudocode for this embedder can be found below in \cref{algo:mean_lsh_embedder}.

\paragraph{Single LSH} \slsh{} is similar to \lsh{} but instead treats the binary vector as a single index into the OOV embedding table. This means similar users/items with the same LSH vector will have the same embedding. Conversely, users/items with different LSH hashes will have completely different embeddings. As with \lsh{}, the projection matrix remains constant, but the OOV embedding table values are updated during training.

\begin{algorithm}[!ht]
    \caption{PyTorch-style pseudocode for the \rbuck{} OOV embedder.}%
    \label{algo:random_embedder}
\begin{minted}[mathescape,
               linenos,
               numbersep=5pt,
               framesep=2mm]{python}
# oov_id: OOV user/item ID
# oov_table: OOV embedding table.
# Each row of the table is an OOV bucket
def rbucket_embed(oov_id, table):
    # b is the number of rows in oov_table
    b = oov_table.size(0)
    # hash_func is a deterministic hash function
    # that returns an integer
    hashed_id = hash_func(row_features)
    # we use the selected bucket's embedding
    return oov_table[hashed_id %
    # the bucket is updated via backpropagation
\end{minted}
\end{algorithm}

\begin{algorithm}[!ht]
    \caption{PyTorch-style pseudocode for the \lsh{} OOV embedder.}%
    \label{algo:mean_lsh_embedder}
\begin{minted}[mathescape,
               linenos,
               numbersep=5pt,
               framesep=2mm]{python}
# row_features: vector of the user/item features.
# oov_table: OOV embedding table.
# Each row of the table is an OOV bucket
def lsh_embed(row_features, oov_table):
    # lsh_hash is a binary vector
    lsh_hash = random_projection(row_features)
    # get col-wise mean of rows where vec is 1
    return oov_table[lsh_hash].mean(axis=1)
    # oov_table is updated via backpropagation
\end{minted}
\end{algorithm}

For all of the embedders, we implement \textit{per-feature} normalization --- we normalize each feature vector individually before concatenating them together. This is done to ensure that the distance between two users/items is not dominated by a single feature. Otherwise, long, dense features (like content embeddings) or lists of categorical features (like watch history) could dominate the similarity computations for OOV embedding methods like KNN\@.

\subsection{OOV Embedder Training}%
\label{subsec:training}
The training procedure does not need to be modified for the untrained OOV embedders (\cref{subsec:untrained_embedders}) --- they can be applied to a pre-trained model. However, trained embedders (\cref{subsec:trained_embedders}) add additional parameters that need to be optimized over OOV users/items. With a time-based inductive dataset split (details in \cref{sec:datasets}), our training set contains only IV values, and the test set contains OOV values. OOV embedders are only used on OOV values so there is no training data for their parameters if only use the training set.
As such, there are two main ways to generate OOV data in training for optimizing our OOV embedders: (1) withhold training data and use it as OOV samples or (2) generate synthetic OOV samples from the training data.

\paragraph{Withholding Data} Withholding training data to use as OOV samples is the simplest method, but it also reduces the amount of data available for training. This also complicates evaluation when benchmarking trained embedders against untrained embedders since the untrained embedders do not have access to the withheld data. Reducing the amount of data available for transductive training worsens transductive performance, violating the criteria defined in \cref{sec:intro}. For this reason, we choose to use synthetic OOV samples. However, the withholding data approach may be useful in production settings where we often cannot afford to maintain a unique embedding table entry for every user/item and may treat low-frequency IDs as OOV values.

\paragraph{Synthetic Data} A simple way to train OOV embedders without affecting existing performance is to generate synthetic OOV samples. For each user/item, we create an OOV version of it that has the same interactions. We then select a subset with ratio $\oovSampleRatio$ of the OOV samples each epoch to use for training. We then perform feature masking, a common augmentation for self-supervised learning~\cite{grace,bgrl}, with mask rate $\fmaskRatio$ on the features of the OOV samples. This ensures that generated samples do not have the exact same features as the input samples. There are three types of OOV interactions: (IV user) $\rightarrow$ (OOV item), (OOV user) $\rightarrow$ (IV item), and (OOV user) $\rightarrow$ (OOV item). We generate each type with equal probability --- although, in practice, this can be tuned to match the expected distribution of OOV interactions in production.

\paragraph{Maintaining Transductive Performance} When training our OOV embedder, our aim is to maintain the performance of the transductive portion of the model. For example, with synthetic training, interactions that only involve one OOV user/item will result in undesirable updates to the main embedding table. To avoid this, we split each epoch into two training steps. In the first step, we train the model on the original training data --- as we normally would in transductive training. There are no OOV values at this point, so it does not affect any trainable parameters in the OOV embedder. In the second step, we freeze the main embedding table weights and train the model on the synthetic OOV samples. The only parameters that can be updated at this step are those of the OOV embedder. We also checkpoint and restore the optimizer state before and after the second step. This ensures that the OOV training does not affect the transductive portion of the model.
\begin{table*}
	\centering
    \resizebox{\linewidth}{!}{
	\begin{tabular}{c|ccccc}
		\toprule
		\textbf{Dataset} & \textbf{IV Users / Items} & \textbf{OOV Users / Items} & \textbf{Mean User / Item Deg.} & \textbf{\# User / Item Cat. Feat.} & \textbf{\# User / Item Float Feat.} \\
		\midrule
		\yelp{}         & 126,379 / 79,238  & 28,140 / 13,078            & 13.74 / 21.92                  & 3 / 7                              & 18 / 6                              \\
		\lfm{}          & 11,962 / 76,152   & 342 / 17,190               & 53.69 / 8.39                   & 1 / 2                              & 45 / 1                              \\
		\ham{}          & 200,749 /  18,871 & 36,961 / 7,024             & 12.76 / 135.71                 & 7 / 12                             & 0 / 3                               \\
		\snap{}         & 74,700 / 30,757   & 6,610 / 3,941              & 24.47 / 59.42 
                & 57 / 339 & 172 / 899                            \\
		\bottomrule
	\end{tabular}
    }
	\caption{Statistics for each of the datasets used in this work. The number of float/dense features counts the number of distinct dense vectors, not the total number of floating point values (e.g., text embeddings count as a single float feature).}
	\label{tab:dset_stats}
	\vspace{-0.3in}
\end{table*}

\section{Datasets}%
\label{sec:datasets}

\begin{figure*}[h!tbp]
  \begin{subfigure}[t]{.24\linewidth}
    \centering
    \includegraphics[width=\linewidth]{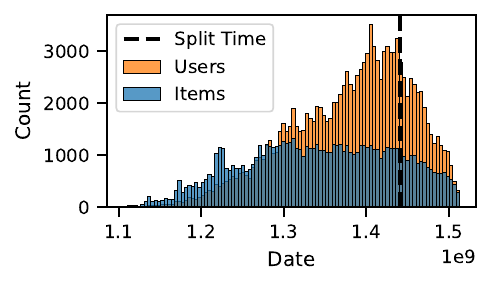}
    \caption{\yelp{}}
  \end{subfigure}
  \hfill
  \begin{subfigure}[t]{.24\linewidth}
    \centering
    \includegraphics[width=\linewidth]{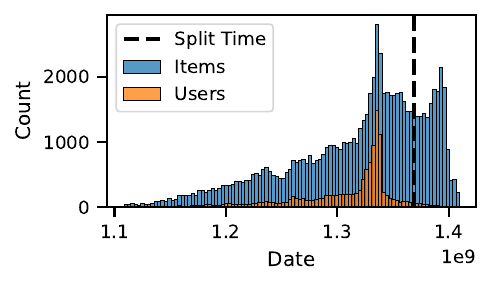}
    \caption{\lfm{}}
  \end{subfigure}%
  \hfill%
  \begin{subfigure}[t]{.24\linewidth}
    \centering
    \includegraphics[width=\linewidth]{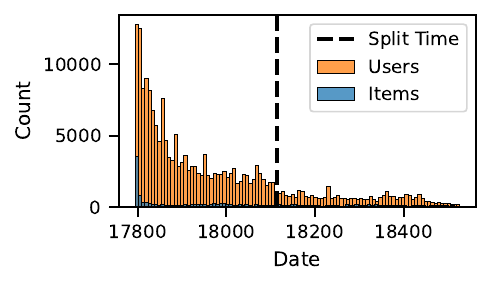}
    \caption{\ham{}}
  \end{subfigure}
  \hfill
  \begin{subfigure}[t]{.24\linewidth}
    \centering
    \includegraphics[width=\linewidth]{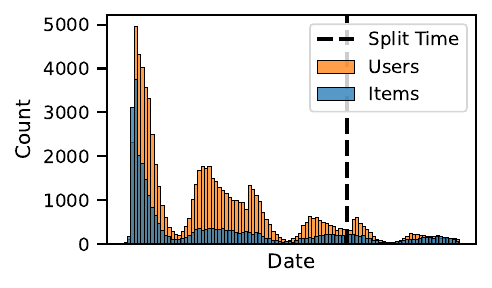}
    \caption{\snap{}}
  \end{subfigure}
  \caption{Visualization of where the inductive split occurs on the datasets. The $x$-axis is the time that the user/item first appeared. Everything to the left of the split time is used for training and validation. The remainder is used for evaluation.}%
  \label{fig:ind_splits}
\end{figure*}

As mentioned in \cref{sec:preliminaries}, following suggestions from recent works~\cite{sun2023take,ji2023critical}, we split the datasets based on a time $t$. We select $t$ for each dataset by computing the first time each user/item appeared. We then select a time $t$ such that 20\% of the users/items are OOV. Formally, select $t$ such that $|\tusers| + |\titems| \approx 0.8 (n+m)$. This results in a naturally different distribution of OOV users compared to OOV items for each of the four datasets. Plots of the relative user/item distributions can be seen in \cref{fig:ind_splits}.

We benchmark various transductive recommendation system methods across four different datasets.
Below, we briefly describe how we processed each of the datasets. Representative statistics for each dataset can be found in \cref{tab:dset_stats}.

\paragraph{Yelp} The \yelp{} dataset consists of user reviews of businesses on Yelp from the 2018 Yelp Dataset Challenge\footnote{\url{https://www.yelp.com/dataset}}. We start with the version of the dataset provided by RecBole~\cite{zhao2021recbole}. We then sample 75\% of the users/items and perform 5-core filtering. We also clean up each feature by removing invalid values, normalizing floating point values, and imputing missing values with scikit-learn~\cite{scikit-learn}. We also remove low-frequency values in categorical features and normalize all strings. Finally, we add text vectors for each business name. We use 300-dimensional GloVe~\cite{pennington2014glove} vectors for this purpose.

\paragraph{LastFM} The \lfm{} dataset~\cite{schedl2016lfm} consists of user/artist interactions on LastFM gathered in 2014. We start with the LastFM-1b version of the dataset provided by RecBole~\cite{zhao2021recbole} and sample 10\% of the users and items. We perform the feature cleaning as with the \yelp{} dataset and add GloVe vectors for each artist's name.

\paragraph{H\&M} The \ham{} dataset consists of user/item purchases on the H\&M website. The raw dataset is taken from the H\&M Kaggle competition\footnote{\url{https://www.kaggle.com/competitions/h-and-m-personalized-fashion-recommendations}} and we sample 30\% of the users/items. We compute GloVe vectors for each item's name and use a pre-trained Vision Transformer~\cite{dosovitskiy2020image} to extract features from each item's image. We also perform the same feature cleaning as with the \yelp{} dataset.

\paragraph{Content} The \snap{} dataset is a proprietary user-item interaction dataset from a large social platform serving hundreds of millions of daily active users. The data is gathered from 5 days of production traffic over users sampled from a single country. We only collect users who are 18 years old and above. The \snap{} dataset has rich user/item features as with many production recommendation systems. Unfortunately, due to the large number of features, we were unable to train any context-aware models with our RecBole-based~\cite{zhao2021recbole} evaluation framework.
\section{Experimental Evaluation}%
\label{sec:experiments}

\begin{table*}[!ht]
	\centering
	\begin{tabular}{r|ccc|ccc|ccc}
		\toprule
		 & \multicolumn{3}{c|}{\ham{}} & \multicolumn{3}{c|}{\lfm{}} & \multicolumn{3}{c}{\yelp{}} \\
		\midrule
		\textbf{OOV Method} & DCNV2             & WideDeep         & xDeepFM          & DCNV2             & WideDeep          & xDeepFM           & DCNV2             & WideDeep          & xDeepFM           \\
		\midrule
		\fdhe{}        & 66.07             & 68.51            & \underline{72.1} & 85.53             & 83.25             & 75.88             & 71.24             & 75.87             & 68.86             \\
		\dhe{}         & 69.09             & 68.55            & \textbf{74.12}   & 86.15             & 84.97             & 59.74             & 70.57             & 67.15             & 71.48             \\
		\zero{}        & \underline{71.06} & 71.21            & 69.06            & 81.57             & 86.57             & 84.75             & 71.16             & 72.43             & 76.04             \\
		\knn{}         & 63.71             & 63.13            & 63.61            & 83.9              & 84.79             & 83.2              & 72.73             & 73.42             & 75.04             \\
		\random{}      & 55.17             & 70.49            & 66.76            & 82.14             & \underline{86.64} & 85.52             & 78.95             & 74.75             & 73.68             \\
		\rbuck{}       & 65.48             & 70.61            & 68.37            & \textbf{87.13}    & 85.79             & 84.24             & \underline{79.88} & 70.97             & 76.04             \\
		\dnn{}         & 63.87             & \underline{71.7} & 71.09            & 86.65             & 84.71             & \underline{86.19} & 76.23             & \underline{77.89} & \underline{82.26} \\
		\slsh{}        & \textbf{73.03}    & \textbf{72.61}   & 70.64            & 86.37             & 79.71             & 85.86             & 78.52             & 76.03             & 80.68             \\
		\meane{}       & 67.72             & 70.72            & 66.12            & 86.49             & 85.79             & 85.16             & 74.9              & \textbf{79.82}    & 73.88             \\
		\lsh{}         & 70.69             & 71.65            & 71.3             & \underline{86.93} & \textbf{86.67}    & \textbf{86.78}    & \textbf{79.96}    & 76.35             & \textbf{82.48}    \\
		\bottomrule
	\end{tabular}
	\caption{OOV user AUC of context-aware methods with different OOV embedding methods. Higher is better. The best-performing method in each column is bolded, and the second-best is underlined. Rows are sorted from lowest mean rank to highest mean rank.}%
	\label{tab:ranking_table}
	\vspace{-0.10in}
\end{table*}

\subsection{Evaluation Details}%
\label{subsec:eval_method}

\begin{figure*}[!htp]
    \centering
    \includegraphics[width=\linewidth]{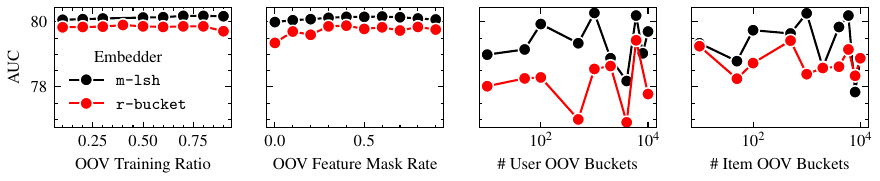}
    \vspace{-0.2in}
    \caption{Sensitivity analysis of different training hyperparameters for \lsh{} and \rbuck{} with WideDeep on \yelp{}. Note that the y-axis range is relatively small and that the x-axis for OOV buckets is on a logarithmic scale.}%
    \label{fig:ablation}
\end{figure*}

\paragraph{Evaluation Metrics} We evaluate the ranking and retrieval models separately. Following conventions from existing work~\cite{dcnv2,dcn,yang2023graph,rendle2012bpr}, we use ndcg@$k$ (where $k$=20) for retrieval models and AUROC for ranking models. In \cref{tab:ranking_table,tab:retrieval_results}, we report the inductive performance of OOV users. It is worth noting that the transductive performance of IV users to IV items remains the same due to how we train the OOV embedding models (see \cref{subsec:training}).

\paragraph{Experimental Details} All models utilize a fork of the RecBole~\cite{zhao2021recbole,zhao2022recbole} framework for experiments, in which we have made extensive modifications to the framework and models to support OOV values and swap between different OOV embedder types. We also added support for filtered evaluation on a subset of users/items. We were very careful to facilitate the easy addition of OOV support to new models. We run all experiments on Google Cloud Platform (GCP). Experiments are conducted on Google Compute Engine instances with NVIDIA Tesla P100 GPUs. The anonymized code, datasets, and hyperparameters for each of our experiments and embedders are available here: \url{https://github.com/snap-research/improving-inductive-oov-recsys}.

\begin{table}[!h]
	\centering
    \resizebox{\columnwidth}{!}{
	\begin{tabular}{r|cc|cc|cc|c}
		\toprule
		\textbf{Dataset} & \multicolumn{2}{c|}{\yelp{}} & \multicolumn{2}{c|}{\lfm{}} &  \multicolumn{2}{c|}{\ham{}} & \multicolumn{1}{c}{\snap{}} \\
		\midrule
		\textbf{Method} & BPR              & DAU         & BPR               & DAU       & BPR              & DAU         & BPR  \\
		\midrule
		\zero{}        & 0.79             & 0.79             & 0.93              & 0.93             & 1.15             & 1.15             & 0.71             \\
		\fdhe{}        & 0.99             & 1.06             & 0.34              & 0.35             & 1.97             & 2.02             & 1.32             \\
		\dhe{}         & 1.12             & 1.11             & 0.59              & 0.40             & 2.09             & 1.96             & 1.39             \\
		\dnn{}         & 2.94             & \textbf{4.43}    & 0.38              & 0.44             & 2.94             & \textbf{3.36}    & 1.87             \\
		\knn{}         & 6.95             & 1.58             & 45.69             & \textbf{4.86}    & 5.23             & 1.67             & \textbf{6.00}    \\
		\random{}      & 4.05             & 0.83             & 15.09             & 0.71             & 2.80             & 1.89             & 0.94             \\
		\rbuck{}       & 9.33             & 1.22             & 41.78             & 0.76             & 6.02             & 1.38             & 1.38             \\
		\slsh{}        & 9.13             & 1.05             & 46.92             & 0.79             & \textbf{6.19}    & 2.02             & 1.05             \\
		\meane{}       & \underline{9.40} & \underline{2.89} & \textbf{48.38}    & 0.12             & 6.15             & 1.94             & \underline{3.74} \\
		\lsh{}         & \textbf{9.49}    & 1.49             & \underline{47.85} & \underline{1.13} & \underline{6.16} & \underline{2.15} & 2.02             \\
		\bottomrule
	\end{tabular}
	}
	\caption{OOV user NDCG@20 of context-free methods with different OOV embedding methods. Higher is better. The best-performing method in each column is bolded and the second-best is underlined.}%
	\label{tab:retrieval_results}
	\vspace{-0.20in}
\end{table}

\subsection{Context-Aware Results}%
\label{subsec:ctx_aware_results}

The OOV user evaluation results of context-aware models are displayed in \cref{tab:ranking_table}. On average, the best-performing OOV embedding method is \lsh{} and the worst is \fdhe{}. Unfortunately, we were unable to train context-aware models on \snap{} (even in the transductive setting) using our RecBole-based framework due to the large number of features and resulting stability issues. We make the following observations:

\paragraph{Context helps OOV embeddings} From \cref{tab:ranking_table}, we can see that incorporating contextual information generally helps OOV embeddings. 
$\sfrac{3}{4}$ of the best-performing OOV embedding models utilize context information. This aligns with our intuition: similar users/items should have similar embeddings. This is true for both context-free and context-aware models. In some cases, like with xDeepFM on \yelp{}, the gap in AU-ROC on OOV users is as large as 6 points --- showing that incorporating feature information in OOV handling can drastically improve an RS's ability to generalize to OOV users/items.

\paragraph{LSH-based solutions perform well} Both \lsh{} and \slsh{} work well for the context-aware models, with one of the two methods performing the best on $\sfrac{6}{9}$ model/dataset combinations, as shown in \cref{tab:ranking_table}. Across the context-aware model experiments, \lsh{} and \slsh{} show a mean improvement of $3.74\%$ and $2.58\%$ over \rbuck{} (a common industry standard\footref{foot:examples}), respectively. They also perform well compared to the next-best method, $\meane{}$, showing respective average improvements of $3.45\%$ and $2.25\%$.

\paragraph{DHE-based solutions perform poorly.} \dhe{} and \fdhe{} are the methods with the lowest average rank across the model/dataset combinations shown in \cref{tab:ranking_table}. Surprisingly, we find that \zero{} generally outperforms both \dhe{} and \fdhe{}. This is likely due to the additional noise the multiple hash inputs introduce to DHE-style models --- a different ID results in a completely different embedding.

\subsection{Context-Free Results}%
\label{subsec:ctx_free_results}

\cref{tab:retrieval_results} shows the NDCG@20 for OOV users of BPR and DirectAU. \lsh{} has the highest mean rank of the different OOV embedding methods. Unlike in the context-aware setting, a relatively large gap exists between different base models on the same dataset. Surprisingly, BPR outperforms DirectAU on OOV users across all three of the datasets. We make the following observations about OOV embedding methods on the context-free models:

\paragraph{Improving context-free OOV performance is difficult} Both BPR and DirectAU exhibit poor performance on most of the datasets, regardless of OOV embedder choice. This shows that it is difficult to encode feature information from OOV IDs in a useful manner for context-free models.

\paragraph{OOV Embedder choice is extremely important} From \cref{tab:retrieval_results}, we can observe that there is a large gap between the best-performing models on each dataset and the worst-performing models for context-free models. This is especially true for BPR on \lfm{}, where there is a 48.04 gap between the best-performing \meane{} embedder and the worst-performing \fdhe{} embedder. \fdhe{} and \dhe{} exhibit similarly poor performance across the four datasets.

\subsection{Sensitivity Analysis}%
\label{subsec:ablation}

As mentioned in \cref{subsec:training}, there are two key hyperparameters for training the models: $\oovSampleRatio{}$ (OOV sampling ratio) and $\fmaskRatio{}$ (feature masking probability). \rbuck{}, \lsh{}, and \slsh{} also assign values to buckets in an embedding table. These methods, therefore, have another hyperparameter $b$, the number of buckets, for each instance of the OOV embedding method. Since we use two instances of each OOV method (one for user IDs and one for item IDs), each model run has two bucket-related hyperparameters: $b_u$ and $b_i$, the number of user/item buckets, respectively.

We conduct a sensitivity analysis on each of the four hyperparameters on \lsh{}, the best-performing OOV embedder, and \rbuck{}, a frequently-used approach in practice\footref{foot:examples}. The results of this analysis are shown in \cref{fig:ablation}. Generally, the performance is not very sensitive to OOV training ratio and feature mask rate hyperparameters; even for the number of user/item OOV buckets, the only performance fluctuates within a relatively small range. We can also observe that \lsh{} outperforms \rbuck{} even under different low training ratios and high feature mask rates. 

\subsection{Recommendations for Practitioners}%
\label{subsec:takeaways}

Based on the results of our experiments in \cref{tab:ranking_table,tab:retrieval_results}, we make the following recommendations for practitioners aiming to improve their performance on OOV users/items:

(1) \textbf{If contextual information (features) is available}, try using \lsh{}. Across our experiments, \lsh{} generally performs the best. An advantage of \lsh{} over \slsh{} is that it results in fewer collisions (see \cref{table:embedder_comparison}). It can also be trivially computed directly on the GPU and efficiently implemented through data structures like PyTorch's~\cite{pytorch} \href{https://pytorch.org/docs/stable/generated/torch.nn.EmbeddingBag.html}{\texttt{EmbeddingBag}}.

(2) \textbf{If no features are available and collisions are not important}, consider using \meane{}. It is extremely cheap to compute and, based on our experiments, is the best-performing untrained OOV embedder. However, all IDs will receive the same embedding, making it particularly problematic for context-free models.

(3) \textbf{If only users \textit{or} items have features}, OOV embedding methods can be mixed. For example, in a dataset with user features but no item features, \lsh{} could be used for users and \meane{} for items. This approach can also be used in any case where the user/item ID distributions are significantly different.

(4) \textbf{Careful caching can greatly speed up OOV training/inference}. Many of the OOV embedding methods described in this work can benefit from caching --- usually for the mapping from ID to bucket(s). For example, caching the LSH vector in \lsh{} and \slsh{} can save a vector-matrix multiplication for each ID. Similarly, caching the $k$-nearest-neighbors for \knn{} can save an ANN index query. This is especially true in cases where features are static or during training, during which the full set of examples is known.

\section{Additional Related Work}
\label{sec:related}
\paragraph{Hashing for Scalable RS} While this work primarily utilizes hashing to support unseen IDs, hashing is often used to improve the scalability of practical RS. One method is via the ``hashing trick''~\cite{moody1988fast,weinberger2009feature}, which reduces the feature space required by categorical features. \citet{zhang2020model} uses two hash functions to select and combine the embeddings of high-frequency items to form the embeddings for low-frequency items. \citet{liu2022monolith} uses cuckoo hashing~\cite{pagh2004cuckoo} on high-frequency IDs to maintain a collision-less hash table. This allows for the continuous eviction of old IDs during online training. \citet{zhang2018efficient} improves the speed of recommendations by using LSH to limit candidate pairs. \citet{ghaemmaghami2022learning} proposes a novel hierarchical-clustering-based approach to hash users/items to encourage collisions between similar IDs, resulting in better performance with fewer buckets. \citet{zhang2016discrete} formulates collaborative filtering as a binary code hashing problem, allowing users/items to be represented with binary embeddings. This allows for improvements in speed and storage efficiency. \citet{tan2020learning} instead utilizes a Graph Neural Network (GNN) to learn the binary hashing function.

\paragraph{Cold-Start RS Methods} A known issue in recommendation systems is the cold-start problem~\cite{lika2014facing}, which is when low-degree users and items receive poorer quality recommendations. In this work, we look specifically at the problem of OOV users/items, which means they occur exactly \textit{zero} times in the training examples. However, cold-start methods often focus on the transductive setting where all the users/items appear at train time (although some may have very few interactions). \citet{vartak2017meta} focuses on the case of OOV items and proposes a meta-learning approach that uses a classifier based on user history to adjust model parameters. \citet{wang2021preference} extends Model-Agnostic Meta-Learning (MAML)~\cite{finn2017model} for improving cold-start recommendation performance. DropoutNet~\cite{volkovs2017dropoutnet} uses input dropout during RS model training to improve the model's generalization to missing features. \citet{lam2008coldstart} proposes a probabilistic approach to handling OOV users on MovieLens~\cite{movielens}.

\paragraph{Cold-Start Graph Methods} Recommendation systems can be formulated as a \textit{link prediction} problem on a bipartite graph, where edges represent interactions between users and items. As such, we also briefly discuss existing literature focused on improving cold-start performance on graph-related tasks. These include both training-based~\cite{hu2022tuneup,zheng2021cold} and augmentation-based~\cite{rong2019dropedge,zhao2022learning} approaches. However, due to model architecture and training differences, these approaches are not straightforward to apply to RS.

\paragraph{NLP OOV Handling} There has also been extensive study of handling OOV values for text tokenization in the field of Natural Language Processing (NLP)~\cite{mielke2021between,bojanowski2017enriching}. Modern DNN models typically use sub-word (e.g., character or byte-level) tokens~\cite{elmo,bert,mielke2021between}, which reduce or eliminate the chance of OOV values. However, word-level tokenizers often have to deal with OOV values, and various approaches have been proposed, including using mean and random vectors~\cite{mielke2021between,lochter2020deep,lochter2022multi}.
\section{Conclusion}
\label{sec:conclusions}

In this work, we explored the inductive setting in recommendation systems, where we focused on finding the best method to handle previously unseen (OOV) values. We evaluated nine different OOV embedder methods that are efficient, model-agnostic, and guaranteed to maintain transductive performance. To the best of our knowledge, this is the first comprehensive empirical study of the performance of various OOV methods for recommendation systems. Our results show that, of the nine methods, the locality-sensitive-hashing-based methods tend to be the most effective in improving inductive performance. Additionally, we augment and re-release three inductive datasets to facilitate future study of inductive performance and OOV methods in recommendation system problems. Furthermore, we derive a set of four recommendations for industrial practitioners to improve their inductive recommendation systems performance and alleviate pain points in dealing with OOV values. We hope this work encourages both academic and industrial researchers to further explore the inductive and OOV settings, considering their immediate practical impact in real-world, production-scale recommendation systems. 

\clearpage
\bibliographystyle{ACM-Reference-Format}
\ifnum\pdfstrcmp{\jobname}{output}=0 %
    \bibliography{bib/vagelis_refs,bib/refs.bib}
\else %
    \bibliography{bib/vagelis_refs,bib/refs.bib}
\fi
\balance%

\end{document}